\documentclass[apjl]{emulateapj}

\usepackage{graphicx}
\usepackage{apjfonts}
\slugcomment{Submitted to ApJ letters}
\shorttitle{Height variation of the vector magnetic field in solar spicules}
\shortauthors{D.\ Orozco Su\'arez, et al.}

\newcommand{\degree}{\ensuremath{^\circ}\/}

\begin{document}

\title{Height variation of the vector magnetic field in solar spicules}

\author{D.\ Orozco Su\'arez\altaffilmark{1,2}, A.\ Asensio Ramos\altaffilmark{1,2}, and J.\ Trujillo Bueno\altaffilmark{1,2,3}}

\email{dorozco@iac.es}

\altaffiltext{1}{Instituto de Astrof\'isica de Canarias, E-38205 La Laguna, Tenerife, Spain}
\altaffiltext{2}{Departamento de Astrof\'isica, Universidad de La Laguna, E-38206 La Laguna, Tenerife, Spain}
\altaffiltext{3}{Consejo Superior de Investigaciones Cient\'ificas, Spain}

\begin{abstract}
Proving the magnetic configuration of solar spicules has hitherto been difficult due to the lack of spatial resolution and image stability during off-limb ground-based observations. We report spectropolarimetric observations of spicules taken in the \ion{He}{1}~1083~nm spectral region with the Tenerife Infrared Polarimeter II at the German Vacuum Tower Telescope of the Observatorio del Teide (Tenerife; Canary Islands; Spain). The data provide the variation with geometrical height of the Stokes I, Q, U, and V profiles whose encoded information allows the determination of the magnetic field vector by means of the HAZEL inversion code. The inferred results show that the average magnetic field strength at the base of solar spicules is about 80 gauss and then it decreases rapidly with height to about 30 gauss at a height of 3000~km above the visible solar surface. 
Moreover, the magnetic field vector is close to vertical at the base of the chromosphere and has mid inclinations (about 50 degree) above 2~Mm height.  
\end{abstract}


  \section{Introduction}
  \label{sec1}

Spicules are thin and very dynamic needle-shaped structures, best seen at the solar limb, whose magnetic properties are not well constrained to date. A detailed study of spicules require very high spatial and temporal resolution observations, being the main reason behind their poor characterization. Modern solar space-based observatories, such as the Japanese satellite Hinode \citep{2007SoPh..243....3K} and, more recently,  NASA's Interface Region Imaging Spectrograph (IRIS; \citealt{2014SoPh..289.2733D}) have improved some of the spicules properties by using fast cadence imaging and spectroscopy, respectively  (e.g., \citealt{2007PASJ...59S.655D,2012ApJ...759...18P,2011NewA...16..296T} and references therein). However, their magnetic configuration is not yet well constrained. 

The magnetic field in spicules has been determined using direct analyses of spectral lines in polarized light, mostly using the \ion{He}{1}~1083.0~nm triplet and the \ion{He}{1}~D$_3$~587.6~nm line, or indirectly by using wave properties  \citep{2007A&A...474..627Z,2008JKAS...41..173K,2011ApJ...733L..15V}. From the interpretation of spectropolarimetric observations, we know that at chromospheric heights, spicules harbor field strengths of the order of 10 gauss and inclinations of about 35\degree\/ with respect to the local vertical, although stronger magnetic fields were also found in spicules \citep{2005ApJ...619L.191T,2005A&A...436..325L,2006ASPC..358..448R}. Later, \cite{2010ApJ...708.1579C} carried out a 45 minute time series in sit-and-stare mode in the \ion{He}{1}~1083.0~nm triplet, with a spatial resolution of about 2\arcsec, and at two different heights above the visible solar surface. They interpreted the data with the HAZEL (from HAnle and Zeeman Light) code \citep{2008ApJ...683..542A} and obtained field strengths up to 50 gauss and field inclinations of 35\degree-40\degree. 

In this Letter, we present observations in the \ion{He}{1}~1030~nm triplet obtained under excellent seeing conditions and report the variations of the magnetic field vector of solar spicules with geometric height. We describe the observations in Section 2. In Section 3 we briefly explain the diagnostic technique, and in Section 4, we present the results and the discussion. 

 \begin{figure}[!t]
\begin{center}
\plotone{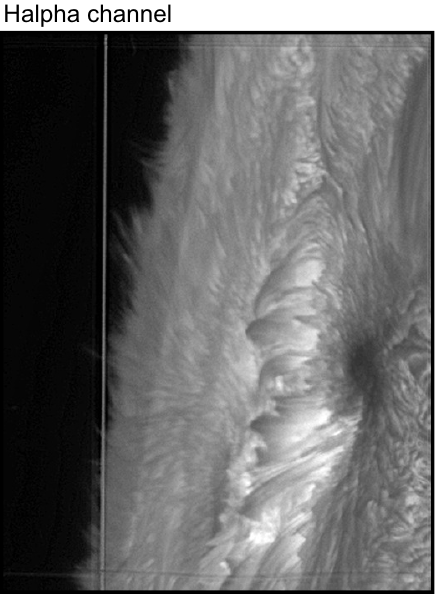}
\caption{Image taken from the H$_\alpha$ slit-jaw camera at the VTT. In the image it can be distinguished a sunspot (right side), where the adaptive optics system of the German VTT was locked, and the target of TIP-II, the spicules. The vertical line represents the spectrograph's slit. In the image it can be seen that the spicules have a dominant orientation.}
\label{fig1}
\end{center}
\end{figure}

\section{Analysis of the observations}
\label{sec2}

The observations were taken on 2013 Apr 23 with the TIP-II instrument \citep{2007ASPC..368..611C} installed at the German VTT of the Observatorio del Teide (Tenerife, Spain).  During the observations, the Kiepenheuer Adaptive Optics System (KAOS, \citealt{2003SPIE.4853..187V}) of the VTT was continuously locked on a sunspot located at the east limb, during good seeing observing conditions. The slit of the spectrograph was situated crossing the solar limb with an angle of $\sim$~45\degree\/ between the slit and limb. Then, an area of about 9\arcsec\/ (25 slit steps with 0\farcs36 step size) was scanned. For each slit position, the TIP-II camera recorded the four Stokes parameters around the 1083.0~nm spectral region with a spectral sampling of 1.1 pm/px. The scanned spectral region contains the chromospheric \ion{He}{1}~1083.0~nm triplet, the photospheric \ion{Si}{1}~1082.70~nm line, and an atmospheric water vapor line at 1083.21~nm. The length of the spectrograph slit is 80\arcsec\/ and the spatial sampling along the slit is 0\farcs17. The slit position of the first scanning step can be seen in the  H$_\alpha$ slit-jaw image (Fig.~\ref{fig1}). The total integration time per slit position was 10 seconds (10 accumulations and 250~ms exposure time per modulation state). We estimate that the spatial resolution could be in the 0\farcs7 -- 1\arcsec\/ range. These data were subjected to the typical data reduction process that includes dark current, flat-field, fringe correction, polarimetric demodulation, and polarimetric cross-talk correction.

\begin{figure*}[!t]
\begin{center}
\plotone{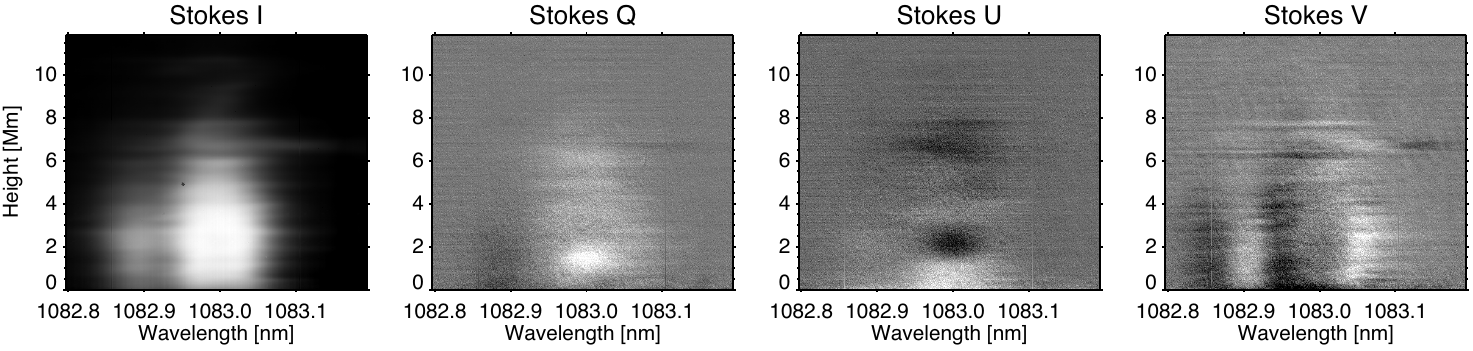}
\caption{Stokes I, Q, U, and V maps representing a single slit position of the full TIP-II map. X-axis represents the wavelength and the Y-axis the height over the solar limb. The Stokes I panel shows the fine structure of the triplet (in emission), with a weak blue component at 1082.90911~nm ($^3$S$_1-^3$P$_0$) separated about 0.12~nm from the other two components at about 1083.02946~nm ($^3$S$_1-^3$P$_1$ and $^3$S$_1-^3$P$_2$), which are indistinguishable. The two components can also be seen in circular polarization. In Stokes Q and U, only the red component can be seen above the noise level although just above the limb  a slight blue component signal in Stokes Q stands above the noise. This tiny blue component is due to the selective absorption mechanism. The positive reference direction for Stokes Q is the parallel to the solar limb.}
\label{fig2}
\end{center}
\end{figure*}

Figure~\ref{fig2} shows an example of TIP-II full Stokes vector observations in the \ion{He}{1} triplet at 1083.0~nm. In particular, we represent a single slit data taken with the TIP-II camera. The first panel shows the variation of the Stokes I profiles of the \ion{He}{1} triplet, which is seen in emission in off-limb observations, with height above the limb. The slit was crossing the spicular material in an inclined angle with respect to the solar limb. Thus, the variations of the intensity with height in Stokes I correspond to spicules, extending up to 8~Mm. Interestingly, spicules can be clearly distinguished in the spectra as a result of to the excellent seeing conditions and high spatial resolution of the data. In the wavelength dimension, Stokes I shows the fine structure of the \ion{He}{1}~1083.0~nm triplet: a weak blue component at 1082.9~nm and the stronger, two red blended components located at about 1083.03~nm.  

In Stokes Q and U, the signals are clearly seen above the noise level. These signals are due to atomic level polarization, that is, population imbalances and quantum interference between the magnetic sub-levels of degenerate atomic levels generated by anisotropic radiation pumping and the Hanle effect. In our data, the positive reference direction for Stokes Q is parallel to the nearest limb. Thus, the presence of signal above the noise in the Stokes U panel readily indicates the presence of a magnetic field inclined with respect to the local vertical direction (which represents the axis of symmetry of the incident radiation) since, according to the Hanle effect theory, the presence of a magnetic field modifies the atomic level polarization. Interestingly, in the data of Stokes U we can see that there is a change in the sign of the profile (from white, positive, to black, negative) indicating that there are changes in the magnetic field azimuth with height in solar spicules. The physics of the Hanle and Zeeman effects in this triplet is  described in the papers by \cite{2002Natur.415..403T}, \cite{2007ApJ...655..642T}, and in the book of \cite{2004ASSL..307.....L}.

Finally, Fig.~\ref{fig2} (right panel) shows that there is a considerable amount of Stokes V signals in the data well above the noise level, i.e., circular polarization profiles dominated by the Zeeman effect, with two lobes of opposite sign. The mere presence of Stokes V signals allow us to determine the magnetic flux (through the Zeeman effect). Given that Stokes Q and U are dominated by atomic level polarization and the Hanle effect and that these signals allow us to determine the field orientation, we can determine the strength of the magnetic field vector rather accurately from the combined interpretation of the Zeeman and Hanle signals. 

\begin{figure*}[http]
\begin{center}
\plotone{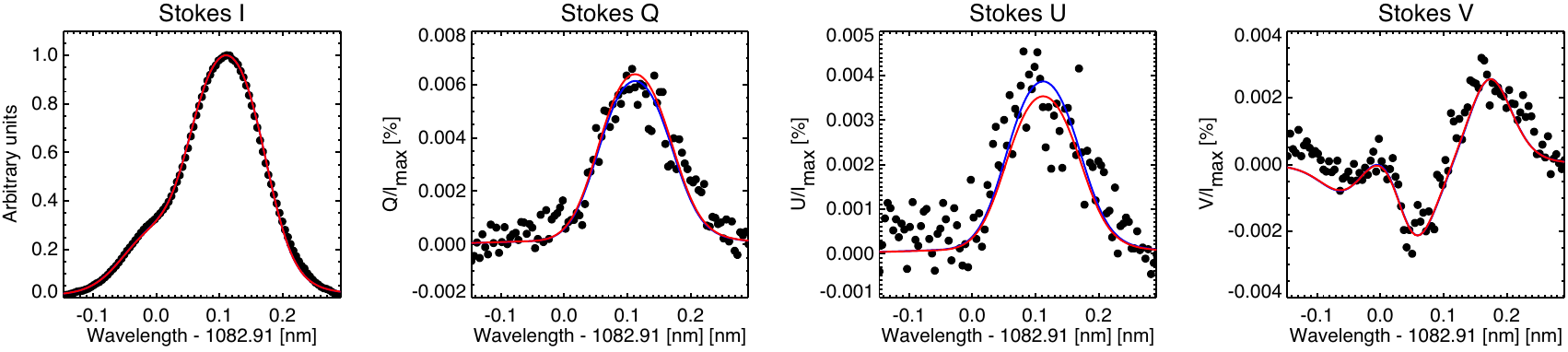}
\caption{Observed (dots) and best-fit (solid line) \ion{He}{1}~1083.0~nm triplet Stokes profiles 
corresponding to a prototypical spicular profile. Stokes I is normalized to unity, while Stokes Q, U, and V are normalized to their Stokes I maximum peak value. Different color codes represent different solutions. The positive reference direction for Stokes Q is the parallel to the solar limb.}
\label{fig3}
\end{center}
\end{figure*}

\section{Inversion of the Stokes profiles}
\label{sec3.2}

In order to interpret the intensity and polarization profiles that we measure in solar spicules we need to assume a suitable model. In our case, we assume a constant-property slab model with a given optical depth $\tau$ \citep{2005ApJ...619L.191T}. We assume that the spicular material of the slab, located at a given height above the limb, is equally illuminated by the unpolarized, limb-darkened photospheric radiation field producing atomic level polarization. We also assume a constant thermal width of the line (modeled with a line damping and a Doppler broadening parameters) and a deterministic magnetic field vector. For computing the Stokes profiles, we take into account the joint action of atomic level polarization and the Hanle and Zeeman effects in the incomplete Paschen-Back effect regime. The slab model takes into account atomic level polarization for all the involved atomic levels and therefore it naturally explains the selective absorption/emission mechanisms we detect in the profiles. In the slab model, the emergent Stokes vector is: 
\begin{equation}
\mathbf{I} = e^{-\mathbf{K}^*\tau}\mathbf{I}_\mathbf{sun} + {\mathbf{K}^{*}}^{-1}\left(1-e^{-\mathbf{K}^*\tau}\right)\mathbf{S},
\end{equation}
where $\mathbf{S}=\epsilon/\eta_\mathrm{I}$ and $\mathbf{K}^*=\mathbf{K}/\eta_\mathrm{I}$, with $\epsilon$, $\eta_\mathrm{I}$, and $\mathbf{K}$ the emission and absorption profiles for Stokes I and the propagation matrix, respectively. In off-limb observations, $\mathbf{I}_{\mathbf{sun} } = 0$. To invert the observed profiles, we use the HAZEL code. The code recovers seven parameters from the inversion of the observed Stokes profiles: the optical depth of the line at the central wavelength of the red blended component $\tau_\mathrm{red}$, the line damping, the Doppler width, the bulk velocity of the plasma, the magnetic strength B, inclination $\theta$, and azimuth $\chi$ of the magnetic field vector.  For reference, the inclination of the magnetic field vector is given with respect to the solar vertical (90 degrees means parallel to the solar surface), while the azimuth of the magnetic field vector has its origin in the line-of-sight direction and is measured counterclockwise (zero degree azimuth and 90 degree inclination corresponds to a magnetic field vector pointing to the observer; see also Fig.~3 in \citealt{2014A&A...566A..46O}). The height above the limb,  which fixes the degree of anisotropy of the incident radiation field, is calculated from the data. A detailed description of the HAZEL code and the underlying physics can be found in the papers by \cite{2002Natur.415..403T}, \cite{2007ApJ...655..642T}, and \cite{2008ApJ...683..542A}. Finally, to improve the signal-to-noise ratio of the observations, the data were down-sampled spectrally and spatially along the slit direction, yielding a final spectral and spatial sampling of 4.4~pm and 0\farcs51, respectively. 

\begin{figure*}[http]
\begin{center}
\plotone{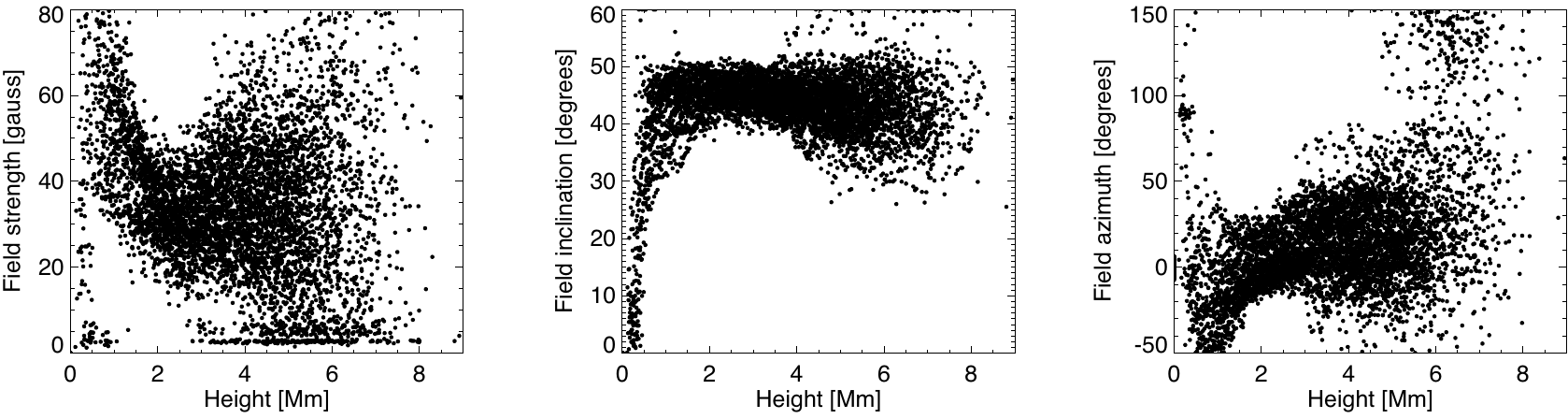}
\caption{Variation of the magnetic field vector (strength, inclination, and azimuth, bottom panels) with height above the solar limb. The results correspond to the solution that is compatible with a magnetic field vector align with the solar spicules, i.e., the vertical solution. The error in the determination of the height above the limb is given by the spatial resolution along the slit. In our case, it is twice the resolution element, i.e., $\pm$740~km. 
}
\label{fig4}
\end{center}
\end{figure*}

\section{Results and discussion: variation of the magnetic field vector with height}
  \label{sec5.3}

An individual fit (solid lines) to prototypical profiles (dots) of the \ion{He}{1}~1083.0~nm triplet is shown in Fig.~\ref{fig3}. The Stokes I profile is normalized to unity and Stokes Q, U, and V to the peak value of Stokes I (before normalization). The fit is, overall, very good. In this particular case, the field strength we obtain from the HAZEL inversion is $B = [54,51]$~gauss, the field inclination $\theta_\mathrm{B}= [42,79]^\mathrm{o}$, which corresponds to a field vector pointing outward from the solar surface, and the field azimuth $\chi_\mathrm{B} = [1,41]^\mathrm{o}$. Note that the inferred vector magnetic field from the analysis of the \ion{He}{1}~1083.0~nm triplet is subjected to the so-called 90\degree ambiguity of the Hanle effect and to the well-known 180\degree azimuth ambiguity (see Sect.~4.2 in \citealt{2014A&A...566A..46O}, \citealt{2015arXiv150103295M} and more references therein). The first one is associated with the Van Vleck angles (54.74\degree and 125.26\degree) and tell us that the magnetic field inclination may have two ambiguous solutions lying at both sides of any of the Van Vleck angles. The two solutions we quoted above represent the 90\degree ambiguous solutions. One of them is compatible with a magnetic field vector rather inclined with respect to the solar vertical and another with more vertical fields. In both cases, the inferred magnetic field strength and its azimuth change. Regarding the 180\degree azimuth ambiguity, in the case of 90\degree scattering geometry, it implies that the solutions $\theta_\mathrm{B}^* = 180^\mathrm{o} - \theta_\mathrm{B}$ with $\chi_\mathrm{B}^* = -\chi_\mathrm{B}$ produce the same polarization signals, so they are ambiguous. In practice, these solutions correspond to a vector magnetic field pointing towards the solar surface. Finally, the inferred optical thickness at the central wavelength of the red component does not change in both solutions and is, in this particular case, $\tau_\mathrm{red} = 1.3$. 

In Fig.~\ref{fig4} we show the variation of the magnetic field strength, its inclination, and azimuth with height above the solar limb. Between the two 90\degree ambiguous solutions, we have chosen the one that is compatible with more vertical magnetic fields. The reason is twofold. First, in the more vertical solution, the inclination of the projection of the magnetic field vector onto the plane of the sky (the plane perpendicular to the line-of-sight direction) is in line with the average inclinations that can be measured in monochromatic observations of spicules, such as that of Fig.~\ref{fig1}. Second, for a long time now, it has been assumed that  spicules are plasma structures standing nearly vertical on the solar surface. Such  assumption has been adopted for their MHD modeling. Hence, we consider that the vertical solution to be the one that makes sense in the current understanding of spicules (e.g., \citealt{1968SoPh....3..367B}). Nevertheless, note that from the point of view of the \ion{He}{1}~1083.0~nm triplet modeling, the two solutions are equally probable. 

The inversion of the data shows that in solar spicules, the magnetic field strength falls rapidly with height, going from 80~gauss close to the visible limb to about 30 gauss at 3~Mm (on average). Above 3~Mm the dispersion increases with an rms of about 30~gauss and a mean of 40~gauss. The high dispersion might be associated with the noise in Stokes V.  These values agree with previous measurements that were in the range of the tens of gauss for solar spicules \citep{2005ApJ...619L.191T,2010ApJ...708.1579C}.

\begin{figure}[http]
\begin{center}
\plotone{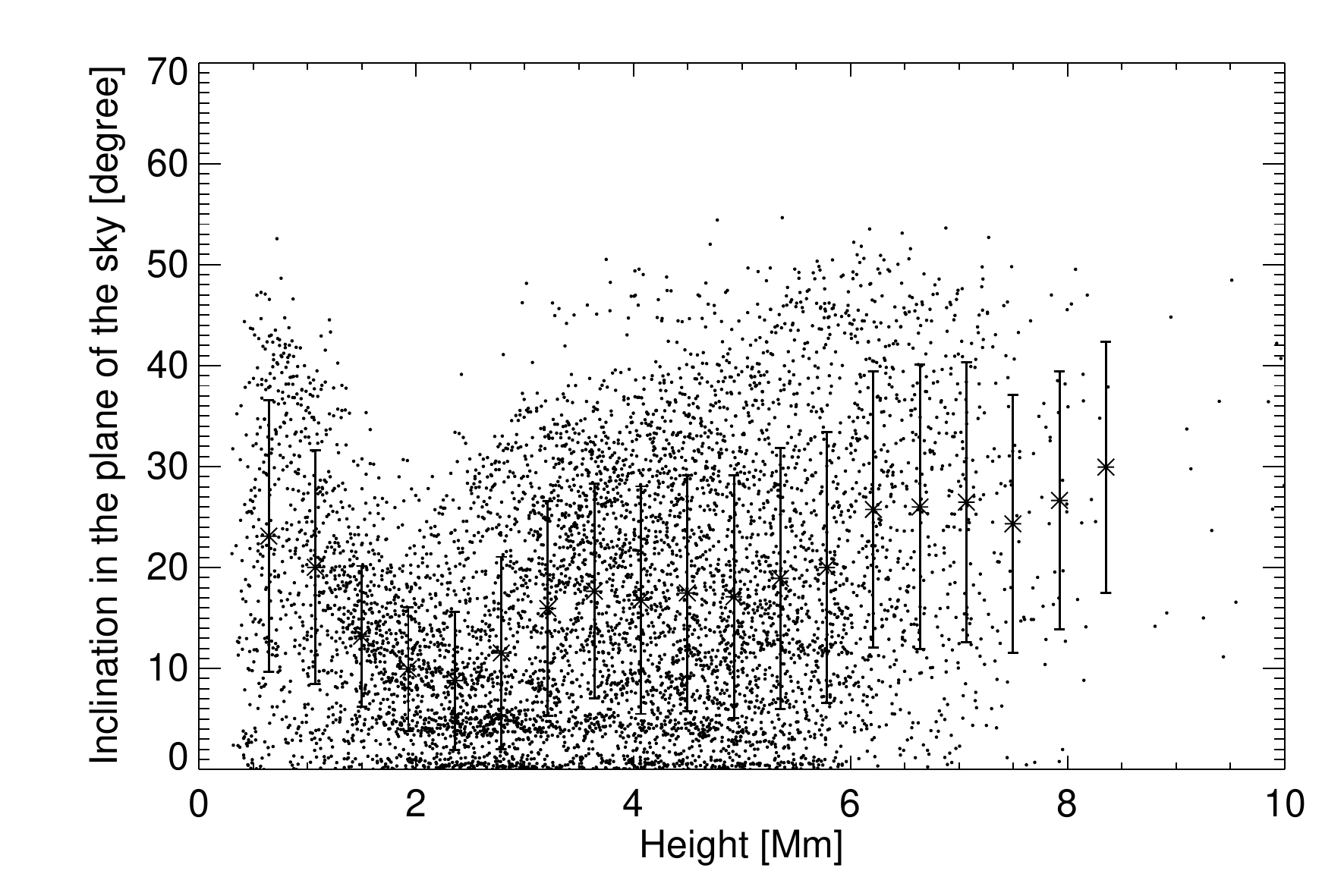}
\caption{Variation of the inclination of the projection of the magnetic field onto the plane perpendicular to the LOS (plane of the sky). The dots represents the inferred values and the asterisks and bars represent the mean and the rms, respectively, for a 500 km bin size in the horizontal axis.}
\label{fig5}
\end{center}
\end{figure}

In the case of the field inclination, at the bottom of the atmosphere, it is close to $0^\mathrm{o}$ (vertical fields) and increases sharply to $\sim$~40\degree\/ above 1.5~Mm. The rms for the inclination also increases, being about $\sim$~5\degree\/ on average, with the rms being less significant than for the field strength. The inclinations we infer agree well with what it can be tentatively derived from monochromatic images. To compare with previous observations, we need to calculate the inclination of the magnetic field projected onto the plane of the sky. The results are shown in Fig.~\ref{fig5}. The mean inclination is about 20\degree\/ with an rms of 11\degree.  

Regarding the field azimuth, it changes from $-50$\degree\/ to  $50$\degree, and the dispersion also increases with height, with an rms of $\sim$~55\degree\/ above 3~Mm. This means that the field vector is pointing toward the observer at almost all heights (azimuths from $-90$\degree\/ to  $90$\degree). Very high in the atmosphere there are positions where the magnetic field points away from the observer. This behavior of the azimuth can be well understood if we have a look at the H$_\alpha$ slit-jaw image of Fig.~\ref{fig1}, where it seems that the spicules contained in our data set have a preferred orientation instead of random orientations. This may be due to the proximity of the active region. 

\section{Conclusions}

In this Letter, we have presented novel spectropolarimetric observations of solar spicules taken in the  \ion{He}{1}~1083.0~nm triplet. We succeeded in partially seeing some of the fine structure of spicules, which is visible up to 8~Mm height. We have detected clear linear polarization signals associated with atomic level polarization and the Hanle effect, while Stokes V is clearly due to the Zeeman effect. The data have allowed us to determine the magnetic field vector variation with height. We find that the field strength in solar spicules decreases from 80 gauss at the visible limb to 30 gauss at about 3000~km, with a large scatter at greater heights,  due to noise and/or averaging along the LOS. No strong variations of the field orientation are found, although the field is clearly more vertical at low heights and then goes close to 40 degree at 1500~km. 
These results represent a step forward in the determination of magnetic fields in chromospheric spicules and should be taken into account in their numerical or theoretical modeling. It would be interesting to record similar observations at higher spatial resolution, e.g., with the GREGOR telescope \citep{2012ASPC..463..365S}, and compare with low resolution results and see if there is any difference in the magnetic field configuration inferred for spicules recorded near active regions,  in the quiet Sun and in coronal holes \citep{1968SoPh....3..367B,2007PASJ...59S.655D,2012ApJ...759...18P}.

\acknowledgments
This work has been supported by the Marie Curie Action: ``International Incoming Fellowships'' of the European Union (FP7-PEOPLE-2012-IIF), Grant Agreement No.~ 327419 and project title MaSFiPro (The three-dimensional Magnetic Structure of solar Filaments and Prominences) and by the Spanish Ministry of Economy and Competitiveness (MINECO) through projects AYA2010-18029 and ESP2013-47349-C6-6-R. The authors gratefully acknowledge the technical expertise and assistance provided by the Spanish Supercomputing Network (Red Espa\~nola de Supercomputaci\'on), as well as the computer resources used: the LaPalma Supercomputer, located at the Instituto de Astrof\'isica de Canarias. A.A.R. also acknowledges financial support through the Ram\'on y Cajal fellowship.


\begin{thebibliography}{}

\bibitem[Asensio Ramos et al.(2008)]{2008ApJ...683..542A} Asensio Ramos, A., Trujillo Bueno, J., \& Landi Degl'Innocenti, E.\ 2008, \apj, 683, 542 

\bibitem[Beckers(1968)]{1968SoPh....3..367B} Beckers, J.~M.\ 1968, \solphys, 3, 367 

\bibitem[Centeno et al.(2010)]{2010ApJ...708.1579C} Centeno, R., Trujillo 
Bueno, J., \& Asensio Ramos, A.\ 2010, \apj, 708, 1579  

\bibitem[Collados et al.(2007)]{2007ASPC..368..611C} Collados, M., Lagg, 
A., D{\'{\i}}az Garcia, J.~J., et al.\ 2007, The Physics of 
Chromospheric Plasmas, 368, 611 

\bibitem[De Pontieu et al.(2007)]{2007PASJ...59S.655D} de Pontieu, B., 
McIntosh, S., Hansteen, V.~H., et al.\ 2007, \pasj, 59, 655 


\bibitem[De Pontieu et al.(2014)]{2014SoPh..289.2733D} De Pontieu, B., 
Title, A.~M., Lemen, J.~R., et al.\ 2014, \solphys, 289, 2733 

\bibitem[Kim et al.(2008)]{2008JKAS...41..173K} Kim, Y.-H., Bong, S.-C., 
Park, Y.-D., et al.\ 2008, Journal of Korean Astronomical Society, 41, 173 

\bibitem[Kosugi et al.(2007)]{2007SoPh..243....3K} Kosugi, T., Matsuzaki, 
K., Sakao, T., et al.\ 2007, \solphys, 243, 3 

\bibitem[Landi Degl'Innocenti 
\& Landolfi(2004)]{2004ASSL..307.....L} Landi Degl'Innocenti, E., \& Landolfi, M.\ 2004, Astrophysics and Space Science Library, 307

\bibitem[L{\'o}pez Ariste 
\& Casini(2005)]{2005A&A...436..325L} L{\'o}pez Ariste, A., \& Casini, R.\ 2005, \aap, 436, 325 

\bibitem[Martinez Gonzalez et al.(2015)]{2015arXiv150103295M} Martinez 
Gonzalez, M.~J., Manso Sainz, R., Asensio Ramos, A., et al.\ 2015, 
arXiv:1501.03295 

\bibitem[Orozco Su{\'a}rez et 
al.(2014)]{2014A&A...566A..46O} Orozco Su{\'a}rez, D., Asensio Ramos, A., \& Trujillo Bueno, J.\ 2014, \aap, 566, AA46 

\bibitem[Pereira et al.(2012)]{2012ApJ...759...18P} Pereira, T.~M.~D., De 
Pontieu, B., \& Carlsson, M.\ 2012, \apj, 759, 18 

\bibitem[Ramelli et al.(2006)]{2006ASPC..358..448R} Ramelli, R., Bianda, 
M., Merenda, L., 
\& Trujillo Bueno, T.\ 2006, Astronomical Society of the Pacific Conference Series, 358, 448 

\bibitem[Schmidt et al.(2012)]{2012ASPC..463..365S} Schmidt, W., von der 
L{\"u}he, O., Volkmer, R., et al.\ 2012, Second ATST-EAST Meeting: Magnetic 
Fields from the Photosphere to the Corona., 463, 365 

\bibitem[Tavabi et al.(2011)]{2011NewA...16..296T} Tavabi, E., Koutchmy, S., \& Ajabshirizadeh, A.\ 2011, NewA, 16, 296 

\bibitem[Trujillo Bueno et al.(2002)]{2002Natur.415..403T} Trujillo Bueno, J., Landi Degl'Innocenti, E., Collados, M., Merenda, L., \& Manso Sainz, R.\ 2002, \nat, 415, 403 

\bibitem[Trujillo Bueno et al.(2005)]{2005ApJ...619L.191T} Trujillo Bueno, 
J., Merenda, L., Centeno, R., Collados, M., 
\& Landi Degl'Innocenti, E.\ 2005, \apjl, 619, L191 

\bibitem[Trujillo Bueno \& Asensio Ramos(2007)]{2007ApJ...655..642T} Trujillo Bueno, J., \& Asensio Ramos, A.\ 2007, \apj, 655, 642 

\bibitem[von der Luehe et al.(2003)]{2003SPIE.4853..187V} von der Luehe, 
O., Soltau, D., Berkefeld, T., \& Schelenz, T.\ 2003, \procspie, 4853, 187 

\bibitem[Verth et al.(2011)]{2011ApJ...733L..15V} Verth, G., Goossens, M., 
\& He, J.-S.\ 2011, \apjl, 733, LL15 

\bibitem[Zaqarashvili et 
al.(2007)]{2007A&A...474..627Z} Zaqarashvili, T.~V., Khutsishvili, E., Kukhianidze, V., \& Ramishvili, G.\ 2007, \aap, 474, 627 


\end{thebibliography}
\end{document}